\documentclass[checkout,showpacs,aps,prb,twocolumn]{revtex4}

\newcommand{\bn}{\begin{enumerate}}
\newcommand{\en}{\end{enumerate}}
\newcommand{\ba}{\begin{eqnarray}}
\newcommand{\ea}{\end{eqnarray}}

\usepackage{graphicx,color}

\newcommand{\be}{\begin{equation}}
\newcommand{\ee}{\end{equation}}
\newcommand{\la}{\langle}

\newcommand{\ra}{\rangle}
\newcommand{\et}{{\it et al. }}
\newcommand{\ete}{{\it et al.}}

\def\prl{{ Phys. Rev. Lett. }}

\def\prb{{ Phys. Rev. B }}

\begin{document}












\title{Ultrafast reduction in exchange interaction by a laser pulse: \\
Alternative path to  femtomagnetism }

\author{G. P. Zhang,$^{1,*}$ Mingqiang Gu,$^{2,1}$ X. S. Wu$^{2}$}
\affiliation{$^1$Department of Physics, Indiana State University, Terre
  Haute, IN 47809, USA\\ 
$^{2}$Laboratory of Solid State
  Microstructures and School of Physics, Nanjing University, Nanjing
  210093, China}

\date{\today}

\begin{abstract}
{ Since the beginning of femtomagnetism, it has been hotly debated how
  an ultrafast laser pulse can demagnetize a sample and switch its
  spins within a few hundred femtoseconds, but no consensus has been
  reached. In this paper, we propose that an ultrafast reduction in
  the exchange interaction by a femtosecond laser pulse is mainly
  responsible for demagnetization and spin switching. The key physics
  is that the dipole selection rule demands two distinctive electron
  configurations for the ground and excited states and consequently
  changes the exchange interaction. Although the exchange interaction
  change is almost instantaneous, its effect on the spin is delayed by
  the finite spin wave propagation. Consistent with the experimental
  observation, the delay becomes longer with a stronger exchange
  interaction pulse. In spin-frustrated systems, the effect of the
  exchange interaction change is even more dramatic, where the spin
  can be directly switched from one direction to the other. Therefore,
  our theory has the potential to explain the essence of major
  observations in rare-earth transition metal compounds for the last
  seven years. Our findings are likely to motivate further research in
  the quest of the origin of femtomagnetism. }
\end{abstract}




 \maketitle

\section{introduction}


In magnetism, exchange interactions among electrons play a central
role in sustaining a long-range magnetic ordering across lattice sites
\cite{stohr,ref15,ref14,ref13,
  ref12,ref11,ref10,ref9,ref8,ref7,ref6,ref5,ref4,ref3,ref2,ref1,
  jpc13}. If the exchange interaction is suddenly quenched, the
consequence would be catastrophic, where the spins could be decoupled
and a magnet could suddenly become nonmagnetic.  In femtomagnetism
\cite{eric, prl00, ourreview,np09,rasingreview}, a laser field excites
electrons out of the Fermi sea and alters the electron configuration
and electron-electron interaction (Coulomb and exchange interactions),
whose change further affects the excitation of the electrons (see
Fig. 1).  Since any pair of electron excitations must obey the dipole
selection rule rigorously, it is conceivable that the effect must be
quite strong.  So far, experimental evidence for the exchange
interaction change has been scant, mostly for the exchange splitting
change in the photoemission spectrum \cite{durr,durr1}.

\begin{figure}
\includegraphics[angle=0,width=8cm]{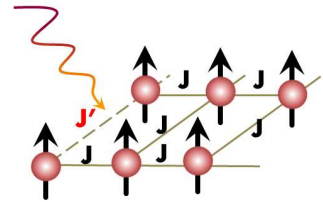}
\caption{Exchange interaction change. 
A laser pulse excites electrons out of the Fermi sea and alters the
electron configuration, and as a result, it changes the exchange
interaction dynamically. 
}
\label{fig1}
\end{figure}

For the Gd(0001) surface states, Lisowski \et \cite{lis} showed that
while the spin polarization is reduced by half, the exchange splitting
is unchanged.  A more recent experiment in gadolinium \cite{weinelt}
shows that the exchange splitting is reduced by 0.2 eV, where the
binding energy increases by 0.1 eV for the minority band and decreases
for the majority band by the same amount.  In $\rm
Gd_{0.55}Sr_{0.45}MnO_3$, Matsubara \et \cite{mat1} showed that it is
possible to switch on the double exchange interaction between ions,
and in $\rm Eu_{1-x}Gd_{x}O$ they \cite{mat2} found that depending on
the doping level of Gd, the exchange interaction can be increased or
decreased.  Up to now, it is unclear whether the exchange change is
the cause or the result of the demagnetization, but as pointed out by
Bigot \et \cite{bigot09}, there is little doubt that the exchange
interaction plays a crucial role here.

Theoretically, the exchange interaction may be temperature dependent
\cite{sz}, but there has been no such study to directly address the
exchange interaction change due to the laser excitation in the time
domain.  The temperature concept is invalid on the femtosecond time scale.
In our first theoretical attempt, we manually changed the exchange
interaction strength \cite{prb98}, and  found that the reduction of
the exchange interaction prolongs the spin relaxation, which was
confirmed experimentally \cite{mathias}. Stamenova and Sanvito
\cite{stamenova} only addressed the dynamical exchange interaction due
to a magnetic field, but their dynamic exchange interaction is still
time-independent since they time-average the exchange coupling
parameter.  Therefore, a theoretical investigation is very appropriate
and important, not only for femtomagnetism, but also for other
research fields.

For instance, ultrafast laser pulses have long been extensively used
to investigate the origin of the high-temperature superconductivity
over two decades \cite{htc,htc1,htc2,htc3,htc4,htc5}. Since the
electron correlation and the electron-phonon coupling act on different
time scales, one can separate them in the time domain.  Similarly,
magnetic and nonmagnetic contributions can be separated as well. By
weakening the exchange interaction, one can suppress the
antiferromagnetic ordering and enhance the nonmagnetic
contribution. This allows one to isolate each effect separately to
examine whether or how the antiferromagnetic coupling is essential to
Cooper pairing in high-temperature superconductors.


In this paper, we theoretically show that a laser pulse is able to
change the exchange interaction dynamically; and as a result, the spin
dynamics (demagnetization and spin switching) is strongly affected.
Due to the optical selection rule, the excited-state configuration
must differ from the ground state, which changes the exchange
interaction. The amount of change depends on the field amplitude and
the helicity of the light, where a much larger change is found for
circularly polarized light.  For a pure ferromagnetic and
antiferromagnetic ordering, for the first time we demonstrate a
delayed response of the spin system with respect to the exchange pulse,
and the delay becomes longer with a smaller exchange interaction,
consistent with the experimental observations. In spin-frustrated
systems, such as rare-earth transition metal compounds, the change is
even more dramatic. For each of three different spin configurations
investigated here, we find that the exchange pulse can change the
course of spin dynamics. Importantly, the spins do not feel the
exchange pulse immediately; instead, after several hundred femtoseconds
when the exchange pulse is over, the main spin change starts, similar
to the experimental results in GdFeCo \cite{stanciu}. By changing the
exchange pulse amplitude, we are able to show that it is indeed
possible that a small change in the exchange amplitude can lead to
dramatically different spin dynamics, but this is only possible in
spin-frustrated systems and when the pulse amplitude is strong
enough. This provides a new clue as to why experimentally changing 
the laser
intensity by little as 0.05 $\rm mJ/cm^2$ could lead to the
helicity-independent switching, a nonthermal all-optical switching.
Therefore, we strongly believe that the exchange interaction change is
an alternative path to femtomagnetism.

This paper is arranged as follows. In Sec. II, we explain how the
exchange interaction is changed during the laser excitation.  Section III
is devoted to the effect of the exchange interaction pulse on the spin
dynamics and switching.  In Sec. IV, we apply our theoretical results
to experiments.  Finally, we conclude the paper in Sec. V.

\section{Exchange interaction change in excited states}

To illustrate how the exchange interaction is changed during the
photoexcitation, we consider a transition from the ground state
$\psi_{gs}=(\phi_a(1)\phi_b(2)-\phi_a(2)\phi_b(1))/\sqrt{2} $ to the
excited state $\psi_{ex}=
(\phi_c(1)\phi_d(2)-\phi_c(2)\phi_d(1))/\sqrt{2}$, where all the
$\phi$ are single-particle wavefunctions. The exchange integrals in
the ground and excited states are $J_{gs}=-\la
\phi_a\phi_b|r_{12}^{-1}|\phi_b\phi_a \ra $ and $J_{ex}=-\la
\phi_c\phi_d|r_{12}^{-1}|\phi_d\phi_c \ra $, respectively, where the
Dirac brackets represent the double integrations over the coordinates
of two electrons and $r_{12}$ is the distance between two
electrons. Here we do not explicitly distinguish the difference
between the exchange interaction and exchange integral, since the
topic has been discussed many times in the literature
\cite{ref15,ref14,ref13,ref12,ref11,ref10,
  ref9,ref8,ref7,ref6,ref5,ref4,ref3,ref2,ref1}.  Usually, these two
exchange integrals are independent of each other, but optical
transitions require that the exchange integral in the excited state
has only 
four independent terms: $-\la
\phi_a\phi_d|r_{12}^{-1}|\phi_d\phi_a \ra $, $-\la
\phi_c\phi_a|r_{12}^{-1}|\phi_a\phi_c \ra $, $-\la
\phi_b\phi_d|r_{12}^{-1}|\phi_d\phi_b \ra $, and $-\la
\phi_c\phi_b|r_{12}^{-1}|\phi_b\phi_c \ra $ (see the appendix A for
details). Since the excited state is much more delocalized, the
exchange interaction is in general smaller.

\begin{figure}
\includegraphics[angle=270,width=7cm]{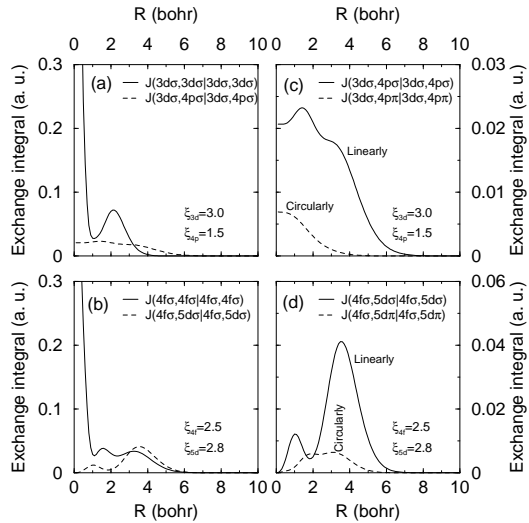}
\caption{Exchange integral as a function of the distance between two
  atoms.  (a) Exchange integral in the ground state
  $J(3d\sigma,3d\sigma|3d\sigma,3d\sigma)$ (solid line) and  in
  the excited state $J(3d\sigma,4p\sigma|3d\sigma,4p\sigma)$ (dashed
  line), mimicking $3d$ transition metals.  (b) Exchange integral in
  the ground state $J(4f\sigma,4f\sigma|4f\sigma,4f\sigma)$ (solid
  line) and  in the excited state
  $J(4f\sigma,5d\sigma|4f\sigma,5d\sigma)$ (dashed line), mimicking
  $4f/5d$ rare earth compounds. (c) Effect of the light helicity on
  the exchange integral for $3d$ systems. The solid line denotes the
  exchange integral for linearly polarized light, while the dashed
  line denotes the integral for circularly polarized light. (d) Effect
  of the light helicity on the exchange integral for $4f/5d$ systems.
  The lines have the same meanings as (c).  The orbital exponents $\xi$ are
  shown in the figure.  }
\label{fig3}
\end{figure}

We can directly evaluate those exchange integrals. Consider two
electrons on two lattice sites separated by a distance $R$. The
ground-state wavefunction is chosen to be the Slater type $3d\sigma$
orbitals localized at two different sites, while the excited-state
wavefunction has one electron in the $3d\sigma$ orbital and the other
promoted to the $4p\sigma$ orbital. This particular choice of the
electron configuration resembles the $3d$ transition metals, excited
with a linearly polarized light. All the exchange integrals are
computed using the Deric program \cite{deric,rue}, with the orbital
exponents $\xi$ and orbital quantum numbers as the input parameters
shown in the figure \cite{freeman,sm}.  Figure \ref{fig3}(a) shows
that $J_{gs}(3d\sigma, 3d\sigma|3d\sigma, 3d\sigma)$ is in general
much larger than $J_{ex}(3d\sigma, 4p\sigma|3d\sigma, 4p\sigma)$ for
most of the distances.

For the rare-earth compounds, the ground state consists of the $4f$
orbital, and the excited state has one electron promoted to the $5d$
orbital. Figure \ref{fig3}(b) shows the same trend for the exchange
integral. However, it has been observed experimentally that the
linearly and circularly polarized lights have different thresholds for
the nonthermal all-optical writing.  We also compute the exchange
interaction change for circularly polarized light. Our finding is very
interesting. Irrespective of whether $3d$ or $4f$ orbitals are
excited, the circularly polarized light reduces the exchange integral
much more. The solid lines in Figs. \ref{fig3}(c) and (d) refer to the
results obtained with the linearly polarized light, while the dashed
lines denote the exchange integrals obtained with the circularly
polarized light. The difference is very clear. The effect in $4f$ is
more pronounced, where the exchange interaction is reduced close to
four-fold. The reason for the reduction is because the circularly
polarized light only excites $\pi$ orbitals. Spatially, there is a
very small overlap with $\sigma$ orbitals, so the exchange integral
becomes smaller. This explains why experimentally the circularly
polarized light is more effective than linearly polarized lights.

As seen from the above discussion, the exchange interaction change
depends on a specific pair of states. Experimentally, a laser pulse
has a narrow photon energy region that naturally select a few
transitions. With a longer laser pulse, the energy window is
narrower. The band structure of a material further limits which states
participate. For the $3d$ transition metals, the transition is mostly
from occupied $3d$ states and unoccupied hybridized $4p$ states. This
means that even for a collective excitation where multiple states
participate, the transitions have the same character.  Our prior study
shows that in fcc Ni, along those high symmetry lines, only two
dominate the initial excitation \cite{prb09}. However, for more
complicated materials, it is necessary to evaluate the exchange
interaction change individually, by taking into account both the
extrinsic (laser field) and intrinsic (material) parameters.

\section{Effects of dynamic  exchange interaction on spin dynamics 
and switching }

Once a laser impinges on a magnetic sample, the laser pulse excites
the electronic systems and changes the exchange interaction. Thus, the
laser effectively introduces a time-dependent {\it exchange
  interaction pulse} \cite{prl08}, \be J(t)=J_0\left (1-A {\rm
  e}^{-(t-T)^2/\tau^2} \right ), \ee where $J_0$ is the static
exchange interaction, $T$ is the time delay of the exchange pulse, and
$\tau$ is the duration time of the pulse, typically on the order of 20
to 100 fs.  $A$ is the dimensionless amplitude changing from 0 (no
laser excitation) to 1 (complete quenching of the exchange
interaction).  To be definitive, we choose $J_0=0.02 $ eV, but one has
to keep in mind that those exchange interaction parameters may differ
a lot for rare-earth compounds. In this study, we have made no attempt
to fine-tune them or distinguish them among different ions, since
there are lots of excellent research done previously in this field
(see, for instance, \cite{wein,ostler}).  Our envelope function is an
inverted Gaussian and ensures that in both the negative and positive
infinite times, the system has the same $J_0$. The exchange drops only
during the narrow time window determined by the duration. In realistic
excitations, the pulse may  not be symmetric in the time domain and may
have a long tail; physically the excited states may live quite a bit
longer than the laser excitation, or the optically excited states might
thermalize into other excited states. As a first step, we assume that
the exchange can be changed only when the electrons transit to the
excited states. This assumption allows us to investigate whether such
a shortest exchange pulse has a lasting impact on the spin dynamics.

In the following, we consider two systems: one-dimensional spin chain
and  a cube of eight spins. It should be pointed out that
due to the Mermin-Wagner theorem \cite{mw}, at any non-zero
temperature, one or two-dimensional isotropic Heisenberg systems can
not be either ferromagnetic or antiferromagnetic. For this reason, our
system is always kept at zero temperature, so the magnetic ordering is
always maintained.

\subsection{One-dimensional spin chain}

To appreciate how the exchange interaction change affects the spin
dynamics, we choose a chain of $N=40$ spins coupled through the
Heisenberg exchange interaction, \be H=- \sum_{ij} J_{ij}(t) {\bf
  S}_i\cdot {\bf S}_j, \ee where ${\bf S}_i$ is the spin operator at
site $i$ and the summation is over the nearest-neighbor sites only.
This Hamiltonian differs from the commonly used Heisenberg model by
the time-dependent exchange interaction \cite{jpc11}. In principle,
$J_{ij}$ should be different for different ions. Wienholdt and
coworkers \cite{wein} even proposed the orbital-resolved exchange
interaction. Our focus is on the qualitative understanding of the
ultrafast magnetization switching among different materials, since
there have been extensive investigations in the literature \cite{spin}
and even the software package is available \cite{oomf}.  We
adopt the semi-classical method that treats the spin as a unit vector.
The dynamic evolution follows the equation of motion \be \frac{d{\bf
    S}_i}{dt}=\sum_{ij}J_{ij}(t) {\bf S}_i \times{\bf S}_j,
\label{eq2}
\ee where the summation is over the nearest neighbors only. Since the
exchange change is almost instantaneous, the spin can feel the
time-dependence of the exchange's change only when the electron is
excited.


\begin{figure}
\includegraphics[angle=0,width=7cm]{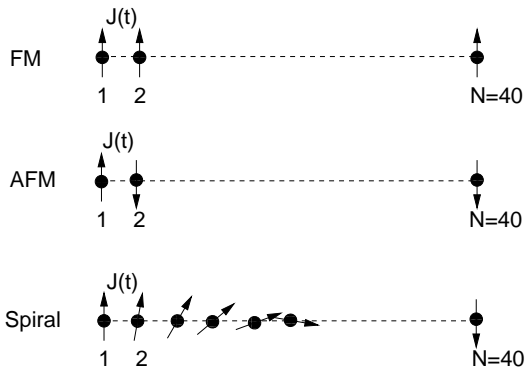}
\caption{Forty-site linear spin chain. Three types of spin configuration
  are considered: Ferromagnetic ordering (FM), antiferromagnetic
  ordering (AFM), and spiral structures (spiral). For all the
  structures except the spiral one, we tilt the first spin 10$^\circ$
  away from the $z$-axis to start the spin dynamics. The spiral spin
  configuration is generated by $S_y(i)=\cos[8\pi (i-1)/N]$ and
  $S_z(i)=\sin[8\pi (i-1)/N]$, where four periods are introduced, $i$
  is the site index, and $N=40$.  }
\label{fig4}
\end{figure}

\begin{figure}
\includegraphics[angle=270,width=7cm]{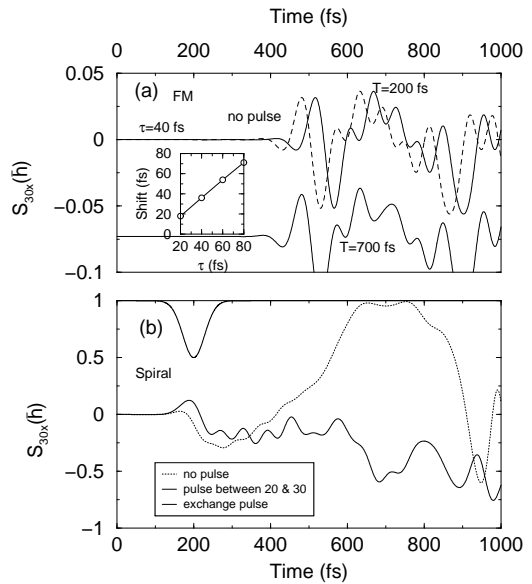}
\caption{ (a) Spin precession with time for the ferromagnetic
  configuration. The solid line represents the $x$-component of spin
  $S_{30x}$ at site 30 with the exchange pulse; the dotted line
  denotes the spin precession without the exchange pulse; and the thin
  line is the spin change for $T=700$ fs (vertically shifted for
  clarity).  Inset: Dependence of the first spin peak time on the
  exchange pulse duration $\tau$ with $T=200$ fs.  Here the exchange
  pulse is applied uniformly to all the sites.  (b) Spin precession as
  a function of time for the spiral configuration.  A dramatic change
  is seen with the exchange pulse applied on sites between 20 to 30.
  The thin solid line is the exchange pulse.  }
\label{fig5}
\end{figure}

 We investigate three different magnetic orderings (see
 Fig. \ref{fig4}): ferromagnetic, antiferromagnetic and spiral
 orderings.  We start with a ferromagnetic configuration with all the
 spins along the $z$-direction initially.  We note in passing that for
 pure ferromagnetic and antiferromagnetic configurations, there is no
 torque exerted on the spins, and an initial tilting is necessary to start
 the spin dynamics. For all the other configurations, this step is
 unnecessary.  To start the dynamics, at $t=0$ we tilt the spin at
 site $i=1$ by 10$^\circ$ from the $z$-axis to $y$-axis and allow the
 spins to precess by themselves for some period of time, since
 experimentally the spin precession is always present even before the
 laser pulse. At $T=200$ fs, we turn on the exchange pulse of
 duration $\tau=40$ fs and amplitude $A=0.5$.  This 200 fs is
 referenced with respect to the initial time $t=0$.  If the initial
 time is shifted to a new time, one has to shift the peak time as
 well.  Figure \ref{fig5}(a) shows that the spin at site ($i=30$)
 $S_{30x}$ precesses with time in the absence (dotted line) and in the
 presence (solid line) of the exchange pulse.  We choose the $x$-component of site $i=30$ because it is representative enough as it is
 far away from the first site.

 Our results show that the spin dynamics with the exchange pulse is
 delayed by 37 fs with respect to that without the pulse.  The
 absolute delay depends on (a) the geometrical distance between the
 excitation site (whose exchange interaction is changed) and the
 location of observation and (b) the nature of the magnetic ordering
 (see below).  We also compute the spin dynamics with $T=700$ fs, where
 the results are similar (see the vertically shifted thin line in
 Fig. \ref{fig5}(a)).  The overall spin precession is shifted to a
 longer time scale, but the details of the dynamics remain the same.
 This is consistent with experimental observations in $3d$ transition
 metals \cite{koopmans,la}.  The reason for this delay can be seen
 directly from Eq. (\ref{eq2}). Once the exchange is reduced, the
 torque on the spin is smaller, and consequently the spin dynamics
 slows down.  This leads to a linear dependence of the delay on the
 pulse duration $\tau$: The longer the pulse is, the longer the delay
 becomes (see the inset in Fig. \ref{fig5}(a) with $T=200$ fs).  This
 conclusion remains the same for the antiferromagnetic ordering and
 the spiral ordering (results are not shown). Other than this time
 delay, the spin dynamics remains the same, but what if not all the
 sites have the same exchange change?

Experimentally, only a small portion of the sample is illuminated by
the laser pulse. We wonder whether this uneven illumination
\cite{kryder} affects the spin dynamics. Nearly 30 years ago, Shieh
and Kryder also invoked such an uneven structure to simulate thermally
driven magnetic switching \cite{kryder}.  We find that for the FM and
AFM chains, the effect is small. However, for the magnetic spiral
structure, the spin precesses very differently from the one without
the exchange pulse. Figure \ref{fig5}(b) shows such a comparison,
where the dotted line represents the results without the exchange
pulse and the solid line represents those with the exchange pulse.
The spin wave reaches site 30 at 100 fs.  We turn on the exchange
pulse at 200 fs only between sites $i=20-30$.  We focus on the spin
change at site 30 as an example. We notice that without the exchange
pulse (see the dotted line), the $x$-component of spin $S_{30x}$
swings from a small negative value, passing through zero, to $1\hbar$,
but with the exchange pulse, $S_{30x}$ mainly remains negative.  The
reason for this is because the initial condition is now changed for
site $i=30$.  By comparing the spiral ordering with the
antiferromagnetic ordering, we notice that the complexity of the local
structure does affect the switching behavior when the exchange
interaction is changed. We also move the peak time of the exchange
pulse to 400 fs, and we reach the same conclusion.  The exchange pulse
affects the dynamics even after the pulse is already gone. Therefore,
it is untrue that the exchange interaction does not play a role beyond
100 fs.  We aim to investigate this further in higher dimensions.

\subsection{A simple cube  of eight spins}

\begin{figure}
\includegraphics[angle=0,width=7cm]{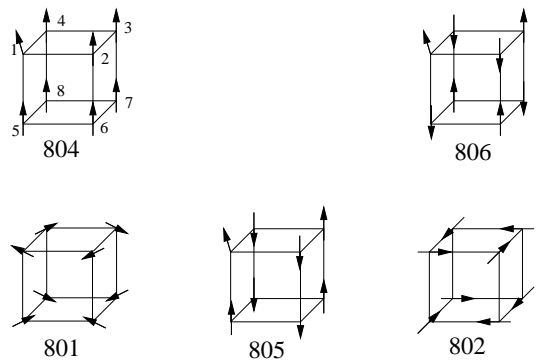}
\caption{Five spin configurations on a cube. The site number
  convention is given on the top left cube. No. 804 is ferromagnetic,
  No. 806 is antiferromagnetic, No. 801 is an in-plane canted spin
  configuration, No. 805 is antiferromagnetic in-plane while
  ferromagnetic inter-plane, and No. 802 has all the nearest
  neighboring spins perpendicular to each other.  These initial
  configurations are used in the real simulation.  }
\label{fig6}
\end{figure}

Up to now, our focus has been on the linear spin chain, where the spin
wave propagates from one site to the next. It is natural to ask what
happens in higher dimensions. Due to the high complexity, we restrict
ourselves to a cube of eight spins and five different spin orderings (see
Fig. \ref{fig6}), with a focus on the main physics.  For larger
systems, the reader is referred to some of the latest publications
\cite{wein}.  We find that despite the diverse magnetic orderings, if
the exchange interaction pulse acts upon all the lattice sites
uniformly and simultaneously, the spin change is also
delayed. Following the above investigation, we consider the cases
where the exchange interaction pulse is only applied upon to sites 1
and 2 (see configuration No. 804 \cite{note1} in Fig. \ref{fig6} for
the numbering convention), while all the other sites have a constant
exchange interaction $J_0$. This approach emphasizes the interaction
change between different spin sites. In our system, we have different
spin configurations between nearest-neighboring sites, where our interest
is on the femtosecond dynamics. Experimentally, GdFeCo samples are
amorphous and exhibit laser-induced ultrafast switching. Building some
inhomogeneity into our model simulates realistic experimental
situations better.

\begin{figure}

\includegraphics[angle=270,width=7cm]{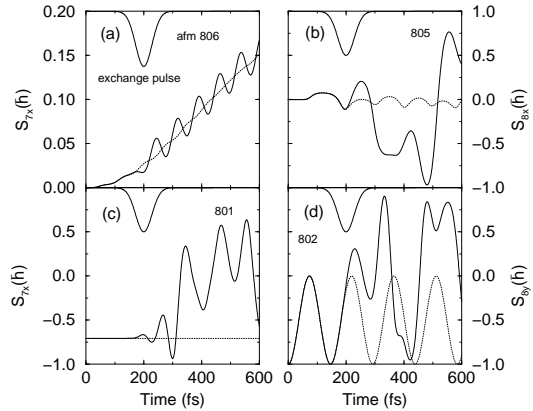}
\caption{ Spin change for four different spin configurations
  (Fig. \ref{fig6}). Here all the exchange pulses peak at $T=200$ fs,
  the exchange pulse amplitude is $A=0.5$, and the duration is 40 fs.
  The dotted line denotes the result without the exchange pulse; and the
  solid line refers to the result obtained with the exchange pulse
  acted upon sites 1 and 2 only. If the exchange interaction acts upon
  all the sites, then a rigid shift of the spin change along the time
  axis is observed.  (a) $S_{7x}$ change with time for the
  antiferromagnetic ordering (No. 806).  The spin on site 1 is tilted
  10$^\circ$ from the $z$-axis initially.  (b) $S_{8x}$ change with
  time for the ordering No. 805. The initial condition is same as
  (a). (c) $S_{7x}$ change with time for the ordering No. 801.  No
  spin is initially tilted.  (d) $S_{8y}$ change with time for the
  ordering No. 802. No spin is initially tilted.  }
\label{fig7}
\end{figure}

With the above consideration in mind, we start to explore the dynamics
in five structures. For FM (No. 804), we find that the overall change
due to the exchange pulse is small (the results are not shown). This
is similar to the linear spin chain. For AFM (No. 806), in
Fig. \ref{fig7}(a) we show how the $x$-component of spin $S_{7x}$ at
site 7 changes with time. This site is chosen since it exhibits a much
more pronounced change.  We notice that without the exchange pulse,
$S_{7x}$ gradually increases with time from 0 fs to 600 fs, with a
very small oscillation (see the dotted line).  When we turn on the
exchange pulse at sites 1 and 2 (see the thin solid line), the
oscillation becomes stronger (see the thick solid line), with its
period determined by the exchange interaction.

However, this is not the case for spin-frustrated orderings. For
instance, configuration No. 805 has an in-plane antiferromagnetic
ordering and an inter-plane ferromagnetic ordering. This structure is
based on the AFM configuration, and the only difference is that we
introduce an inter-plane antiferromagnetic coupling.  Without the
exchange pulse, we find that the $x$-component of spin at site 8 shows
a periodic oscillation from 0 fs to 600 fs (see the dotted line in
Fig. \ref{fig7}(b)). If we turn on the exchange pulse at 200 fs, the
overall change in $S_{8x}$ begins after 200 fs (see the solid line in
Fig. \ref{fig7}(b)). There is no big change before 200 fs, even though
the exchange pulse is already turned on. In other words, there is a
clear delay between the exchange pulse and the spin response. After
300 fs when the exchange pulse returns to $J_0$, {\it the main change
  starts} and $S_{8x}$ swings from $-\hbar$ to $0.75\hbar$. Consistent
with our above findings, this is again opposite to the common belief
that the spin dynamics due to the exchange interaction ceases to occur
after the typical time of the exchange interaction and all the
subsequent spin relaxation results from the phonon.

For other two configurations (Nos. 801 and 802), we reach the same
conclusion. Figure \ref{fig7}(c) shows that without the exchange
pulse, $S_{7x}$ has no clear change, but when the exchange pulse is
introduced, the change starts (see the solid line); the major
change occurs around 350 fs, after which the exchange pulse is completely
restored to $J_0$. In Fig. \ref{fig7}(d) we show the results for
configuration No. 802. Without the exchange pulse, $S_{8y}$ already
oscillates periodically with time, but with the exchange pulse, we see
that the oscillation becomes aperiodic, with the extreme values of
$-1\hbar$ and $1\hbar$. There is also a delay between the exchange
pulse and the major spin change. This is the main finding of our
present study.

\section{Applications to experiments}

\subsection{Delay between the exchange pulse and spin dynamics}

The delay between the exchange pulse and spin dynamics has important
consequences for experiments.  Theoretically, this delay is expected
since the spin wave needs time to propagate from one site to next.
For a normal uniform magnetic ordering such as FM and AFM, the
collective spin excitation can survive the exchange interaction
change.  Physically, they have a larger spin stiffness. This effect is
detectable.  Experimentally, it has been shown that when the laser
intensity increase, the demagnetization time becomes longer
\cite{koopmans,la}.  Koopmans \et \cite{koopmans} showed that
increasing intensity slows down demagnetization in Co and Ni
samples. The net change is 80 fs for both Ni and Co (Ni: from 140 to
220 fs; Co: 160 fs to 240 fs). However, the signal shape remains the
same.

\begin{figure}
\includegraphics[angle=270,width=6cm]{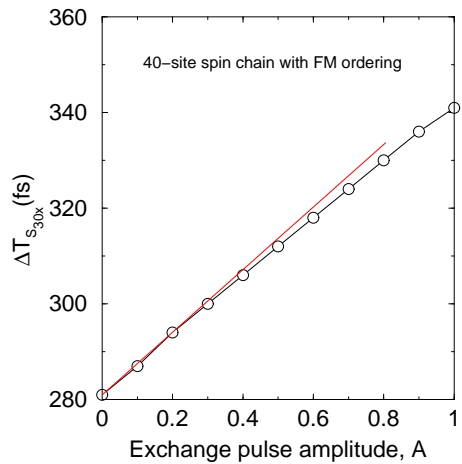}
\caption{ Demagnetization time $\Delta T$ for the spin at site 30 as a
  function of the exchange pulse amplitude.  $\Delta T_{S_{30x}}$ is
  defined as the difference between the first peak time of $S_{30x}$
  and the exchange peak time $T$.  $\Delta T$ increases with $A$, but
  the dependence is not exactly linear; for reference the solid line
  is drawn as the exact linear dependence.  This dependence is similar
  to the experimental observation.  }
\label{fig7.5}
\end{figure}

 Our results suggest that this may be due to the exchange interaction
 reduction induced by the laser field instead of phonons, as shown by
 Lefkidis and H\"ubner \cite{lefkidismmm}. This can be directly seen
 from Eq. (\ref{eq2}), where a small exchange interaction leads to a
 weaker torque on the spin and prolongs the dynamics. Figure
 \ref{fig7.5} shows an example in the ferromagnetic 40-site spin
 chain, where the demagnetization time $\Delta T$ at site 30 increases
 almost linearly with the exchange pulse amplitude $A$. Here the
 demagnetization time $\Delta T$ is defined as the difference between
 the first peak time of the $S_{30x}$ and the exchange pulse peak time
 $T$.  The overall shape does not change strongly with the exchange
 pulse amplitude, similar to the experimental observation.  Second,
 this effect must depend on the sample, since exchange interactions
 and magnetic spin moments are different in different samples.  In 
 Co/Pd multilayer films, Vodungbo \et\cite{noslow} did not detect an
 obvious relation between the pump fluence and the demagnetization
 time.  This finding is very important, as it casts doubt on the
 phonon mechanism (see below for more).

For GdFeCo, the situation is more complicated and more interesting.
Experimentally, Stanciu \et\cite{stanciu} showed that even though
their laser pulse is only 40 fs long, the spin switch occurs on a
picosecond time scale. There is a long delay between the laser initial
excitation and the actual spin switching. As seen from
Fig. \ref{fig7}, the exchange pulse does not act on the spin
immediately; there is a long delay before the actual spin starts to
change. For GdFeCo, the exchange interaction is much weaker, on the
order of 0.01 eV or smaller, so the delay can be much longer. Alebrand
\et \cite{alebrand} found that the helicity information stored in
GdFeCo lasts at least some picoseconds after optical excitation. The
information carrier must not be phononic, since lower temperature
works better for switching as Hohlfeld showed \cite{hohlfeld}.  This
issue has been discussed extensively in our prior study \cite{jpc13}.

Khorsand \et \cite{khor} showed that the threshold fluence for optical
switching is independent of the helicity of the laser pulse, but this
result has not yet been reproduced by other groups. Theoretically, we
find that in general the exchange interaction is reduced more with 
circular polarized light than with  linearly polarized light, so that we
expect  it is easier to switch spins.  With a strong laser,
regardless of its helicity, an even larger reduction in exchange
interaction is expected.

 Experimentally, Ostler \et \cite{ostler} reported an intriguing
 result in GdFeCo that a minor increase in the laser intensity could
 change the spin switching mechanism from a helicity-dependent to
 helicity-independent switch.  Before we compare with the experimental
 results, we would like to resolve a conceptual confusion.  Ostler \et
 \cite{ostler} reported that the ``heating'' alone can result in a
 deterministic reversal. However, their ``heating'' is slightly a
 misnomer since it is not really heating, but instead it refers to
 helicity-independent switching.  Therefore, we suggest that
 helicity-dependent or helicity-independent switching be used to avoid
 unnecessary conceptual confusions in the literature.  
Second, in
 their supplementary material \cite{ostler}, we notice an important
 experimental detail \cite{jpc13}: Increasing the laser intensity from
 2.30 $\rm mJ/cm^2$ to 2.25 $\rm mJ/cm^2$ leads to a transition from a
 helicity-dependent to helicity-independent switching.  Such a
 dramatic switch difference is interesting and puzzling.

\begin{figure}

\includegraphics[angle=270,width=7cm]{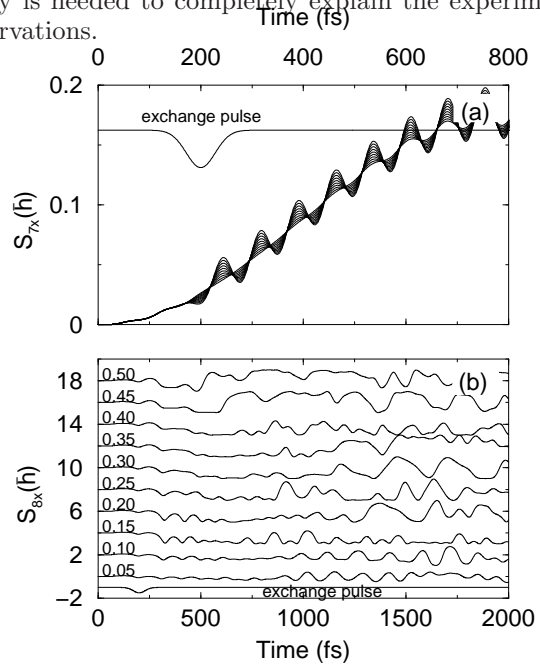}

\caption{ Intensity dependence of the spin switch. (a) For regular AFM
  ordering, the exchange pulse amplitude has no significant effect on
  the spin change. (b) For a frustrated-spin ordering (configuration
  number 805, see Fig. \ref{fig6}), when the amplitude is higher, even
  a small change in the exchange amplitude leads to a dramatic change
  in spin dynamics. This is consistent with the experimental
  observation.  The numbers on each curve are the exchange pulse
  amplitude.  All the curves are shifted vertically for easy viewing.
}
\label{fig8}
\end{figure}

Theoretically, we compute the spin moment change as a function of
various exchange pulse amplitudes $A$ for both AFM (No. 806) and
spin-frustrated orderings (No. 805). We change $A$ from 0.05 to
0.50. The exchange pulse has the same duration and delay as
Fig. \ref{fig7}.  Figure \ref{fig8}(a) shows the results for AFM
ordering. We see that the effect on the spin change is very minor,
only with the oscillation amplitude increased. However, for the
spin-frustrated case (No. 806), we see a big difference. Figure
\ref{fig8}(b) shows that when the exchange amplitude is small, the
change is gradual, but when $A$ is above 0.40,  every 0.05 increase
in the exchange interaction leads to very different spin dynamics.
Once $A$ is above 0.3, the spin switching from the negative maximum to
the positive maximum occurs earlier for $A=0.5$ than for $A=0.3$,
though both the spins reach the first negative maximum at the same
time.  While it is true that our model is still very different from
the real experimental situation, qualitatively our result reveals the
possibility to induce a dramatic change using a strong laser
pulse. Additional study is needed to completely explain the
experimental observations.

\subsection{Phonon controversy}

Many researchers strongly argue that slower demagnetization with a
stronger laser is due to the phonon involvement \cite{koopmans,
  fan,fan1} through the Elliot-Yafet (EY) effect
\cite{elliott,yafet}. But Vodungbo \et \cite{noslow} failed to find
any significant change in the demagnetization. Does this indicate no
phonon involvement?  Theoretically, we find this unsatisfactory.
However, the main evidence against the EY mechanism is from the EY
theory itself.  Regardless of whether it is suitable for
femtomagnetism or not, the theory predicts that the spin relaxation
time $T_1$ is inversely proportional to the number of phonons through
the  equation,\begin{widetext}
 \be \frac{1}{T_1} \propto \int \int dk
dk'G_{k,k'} [n_q\delta (E_{k\downarrow}- E_{k'\uparrow}+\hbar\omega_q)
  +(n_{-q}+1)\delta (E_{k\downarrow}- E_{k'\uparrow}-\hbar\omega_{-q})
] \ee \end{widetext}
where the integrations are over the crystal momentum index
$k(k')$, $G_{k,k'}$ is the matrix element for the spin flip (see
Eq. (18.4) of Yafet's theory \cite{yafet} for details), $n_q$ is the
number of phonons with phonon momentum $q$, $\hbar\omega_q$ is the
phonon energy, and $E_{k\uparrow(\downarrow)}$ is the spin up(down) band
energy.  Since the number of the phonons increases with the laser
intensity naturally, the demagnetization time would become shorter if
the entire process were due to  phonons.  This is just opposite to
what is observed experimentally.

\begin{figure}
\includegraphics[angle=0,width=7cm]{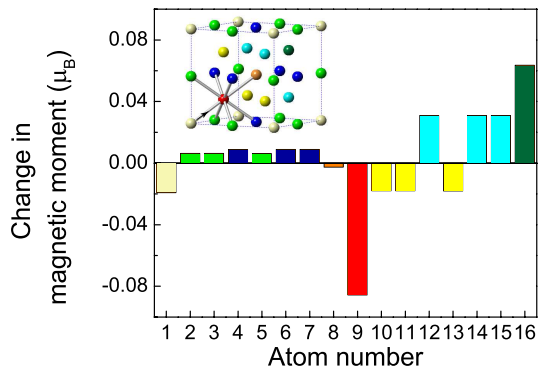}
\caption{Effect of the atom vibration on the spin moment change in bcc
  Fe. A cell of $2\times2\times2$ is chosen. We move atom 1 (with
  arrow) by 0.1 $\rm \AA$ along the diagonal direction and compute the
  spin moment change using the density functional theory.  The largest
  spin moment change occurs for two atoms close to the first atom (see
  the red ($-0.08~ \rm \mu_B$) and dark green ($+0.05~\rm \mu_B$)
  bars). It is obvious that the spin moment change is small. 
}
\label{fig9}
\end{figure}

\newcommand{\ik}{i{\bf k}} To this end, it is clear that the EY
mechanism contradicts the experimental results, but we wonder how much
the spin moment is changed due to the lattice vibration. This has
never been done before.  We take bcc Fe as an example. We construct a
$2\times2\times2$ supercell (see the inset in Fig. \ref{fig9}), which
has 16 Fe atoms in the cell.  Our method is based on the linearized
augmented planewave method as implemented in the WIEN2k code
\cite{wien,prb09}. We use the GGA functional and the $RK_{max}$ is
7.0. The total number of $k$ points is 1000.  We self-consistently
solve the Kohn-Sham equation \be
[-\nabla^2+V_{Ne}+V_{ee}+V_{xc}^\sigma]\psi_{\ik}^\sigma(r)=E_{\ik}^\sigma
\psi_{\ik}^\sigma (r),\label{ks} \ee where the terms on the left-hand
side are the kinetic energy, electron-nuclear attraction, Coulomb and
exchange interactions, respectively.  $\psi_{\ik}^\sigma(r)$ is the
Bloch wavefunction of band $i$ at crystal momentum ${\bf k}$ with spin
$\sigma$, and $E_{\ik}^\sigma$ is the band energy.

To compute the effect of the lattice vibration, we displace one Fe
atom by $0.1\rm \AA$ along the diagonal direction of the cube (see the
inset again). The total energy change due to the Fe displacement is
0.481 eV, already over 10 times higher than the typical phonon energy.
Figure \ref{fig9} shows the atom-resolved spin moment change, with the
same colors representing equivalent spin changes.  The results are
insightful. Only two atoms have a large spin moment change: the spin
moment on No. 9 is reduced by -0.08 $\rm \mu_B$, while that on No. 16
is increased by 0.05 $\rm \mu_B$.  Both atoms are close to the
displaced atom No. 1. Changes on other atoms are below 0.02 $\rm
\mu_B$. These changes are only 3.6\%, in comparison to the  pristine
spin moment of 2.2 $\rm \mu_B$.  Therefore, a simple lattice vibration
only leads to a very small spin moment change \cite{sch}, which is not
enough to explain the experimental findings.


\section{Conclusion}

We have proposed the laser-induced exchange interaction change as an
alternative path to femtomagnetism.  We show that in general the
exchange interaction is reduced for the excited states and the
reduction is helicity-dependent, with a larger reduction found with
 circularly polarized light than with  linearly polarized
light.  Second, we introduce an exchange interaction pulse and
investigate the spin dynamics in two systems with three magnetic
ordering (FM, AFM and spin-frustrated). For pure FM and AFM, we find a
clear delay of the spin dynamics with respect to the exchange pulse,
and the delay becomes longer with a stronger pulse, in agreement with
the experimental results in $3d$ transition metals.  A dramatic spin
change is found in spin-frustrated systems and is also delayed by
several hundred femtoseconds, which is consistent with the observation
in GdFeCo. Finally, we examine the spin change with the exchange pulse
amplitude. We find that for a weak pulse, the change is
small. However, if the pulse is very strong such as $A=0.4$, a change
as small as 0.05 can potentially lead to a strong change. This nicely
explains why experimentally \cite{ostler} only a small increase in the
laser intensity by 0.05 $\rm mJ/cm^2$ could induce a
helicity-independent switch. We believe that it is the exchange
interaction reduction that plays a role here. We also explain why the
phonon effect from the EY theory does not play a critical role here.
Our first-principles calculation shows that even moving one iron atom
by 0.1 $\rm \AA$ along the lattice diagonal, the spin moment is only
reduced by 0.1 $\rm \mu_B$ out of 2.2 $\rm \mu_B$.  Our new mechanism
is expected to inspire new theoretical
\cite{jpc13,ourreview,prl00,np09, prl08,
  lefkidismmm,fan,fan1,prb09,sch} and experimental
investigations \cite{eric,rasingreview,durr1,lis,weinelt,mat1,
  bigot09,mathias,ostler,alebrand,khor,vah1,radu}.

\acknowledgments We would like to thank Prof. Thomas F. George
(University of Missouri-St. Louis) for his interest in our research
and carefully proof-reading our manuscript.  This work was supported
by the U.S. Department of Energy under Contract No.
DE-FG02-06ER46304.  GPZ appreciates help from Prof.  Hagstrom (Indiana
University) with the DERIC program.  Part of the work was done on
Indiana State University's Quantum cluster (funded by the
U. S. Department of Energy) and high-performance computers.  This
research used resources of the National Energy Research Scientific
Computing Center, which is supported by the Office of Science of the
U.S.  Department of Energy under Contract No. DE-AC02-05CH11231. Our
calculations also used resources of the Argonne Leadership Computing
Facility at Argonne National Laboratory, which is supported by the
Office of Science of the U.S. Department of Energy under Contract No.
DE-AC02-06CH11357.

$^*$gpzhang@indstate.edu

\appendix

\section{Special exchange interaction terms during the optical excitation}

To see why the optical transition imposes a restriction on the
electron configuration, we compute the transition matrix elements $\la
\psi_{ex} |D|\psi_{gs} \ra$ for a transition from the ground state
$\psi_{gs}=(\phi_a(1)\phi_b(2)-\phi_a(2)\phi_b(1))/\sqrt{2} $ to
excited state $\psi_{ex}=
(\phi_c(1)\phi_d(2)-\phi_c(2)\phi_d(1))/\sqrt{2}$. A straightforward
calculation yields \begin{widetext}
\be \la \psi_{ex} |D|\psi_{gs} \ra = \la \phi_c
|D_1|\phi_a\ra \delta_{db} + \la \phi_d |D_1|\phi_b\ra \delta_{ca} -
\la \phi_c |D_1|\phi_b\ra \delta_{da} - \la \phi_d |D_1|\phi_a\ra
\delta_{cb}.  \ee \end{widetext}
These delta functions restrict the transition to
occur only with one single-particle wavefunction changed.  As a
result, the exchange integral in the excited state  has only four
independent terms: $-\la \phi_a\phi_d|r_{12}^{-1}|\phi_d\phi_a \ra $,
$-\la \phi_c\phi_a|r_{12}^{-1}|\phi_a\phi_c \ra $, $-\la
\phi_b\phi_d|r_{12}^{-1}|\phi_d\phi_b \ra $, and $-\la
\phi_c\phi_b|r_{12}^{-1}|\phi_b\phi_c \ra $. All the other ones are
forbidden.

\section{Electron-electron interaction change in Particle-in-a-box (PIB)}

To appreciate the change in electron-electron interactions, we compute
the value of Coulomb and exchange interactions in PIB. 
In this model, the potential is zero if the electron position $x$ is
within $[0,L]$, and is infinite if otherwise. The eigenfunction is
well known, \be \phi_n=\sqrt{\frac{2}{L}} \sin (\frac{n\pi x}{L}), \ee
where $n$ is the state index. To overcome the singularity in the
Coulomb potential, we introduce a small broadening $\delta=0.01$ (in
the unit of $L$). This broadening affects the absolute value, but not the qualitative change of our results, which is
verified by using different values of $\delta$.  Our integrals have
the following form: \begin{widetext}
\be U(ab|cd)=\int_{0}^{L}\int_0^L dx_1 dx_2
\phi_a^*(x_1)\phi_b^*(x_2)
\frac{e^2}{|x_1-x_2|+\delta}\phi_c(x_2)\phi_d(x_1). \ee \end{widetext}
We compute
both the Coulomb and exchange interactions between two electrons
numerically using the grid mesh method with a  grid mesh size of
$0.001L$.

\begin{figure}
\includegraphics[angle=270,width=6cm]{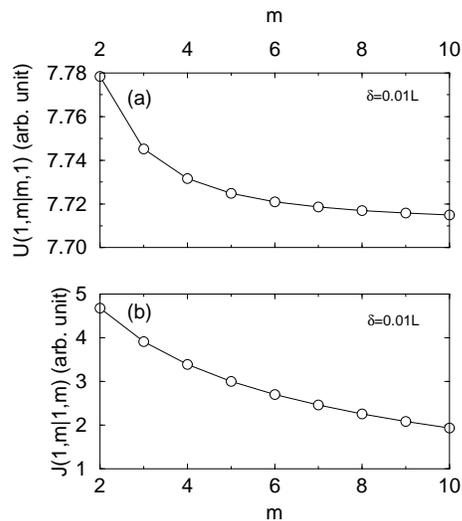}
\caption{Particle in a box.  (a) Coulomb interaction as a function of
  $m$. (b) Exchange interaction as a function of $m$. The integration
  is done over a line segment $[0,L]$.  A small broadening
  $\delta=0.01L$ is used.  }
\label{fig2}
\end{figure}

Since for PIB the states are
not degenerate, for the triplet states, electrons must occupy
different quantum states $(n,m)$, where $n$ refers to the quantum
number for electron 1 and $m$ for electron 2.  We consider that
initially the electrons are in the ground state (1,2). After laser
excitation, the electron configuration becomes $(1,m)$, where $m$ is
larger than 2. Then we compute the Coulomb and exchange matrix
elements for this configuration, $U(1m|m1)$ and $J(1m|1m)$, as a
function of $m$. The results are shown in Fig. \ref{fig2} (a). As $m$
increases, both the Coulomb and exchange interactions decrease. This
is consistent with our above qualitative discussion since the
wavefunction is more diffusive for the excited states and the
integration is smaller. The reduction in the exchange interaction is
much larger than that in the Coulomb interaction. This is expected
since the exchange interaction depends on the product of two different
wavefunctions, or exchange charge, which introduces many more
oscillations to the integration and is very sensitive to the node
change in the wavefunction. By contrast, for the Coulomb interaction,
it is the wavefunction squared, or charge density, that enters the
integration. This reduces the oscillatory component of the
wavefunction, while the nonoscillatory part sets the base value for
the Coulomb interaction.

\begin{thebibliography}{99}


\bibitem{stohr}J. St\"ohr and H. C. Siegmann, {\it Magnetism: From
  Fundamentals to Nanoscale Dynamics}, Springer-Verlag, Berlin,
  (2006).




\bibitem{ref15}J. C. Slater, Phys. Rev. {\bf 49}, 537 (1936).


\bibitem{ref14}H. Brooks, Phys. Rev. {\bf 58}, 909 (1940).



\bibitem{ref13} C. Herring and C. Kittel, Phys. Rev. {\bf 81}, 869
  (1951). 



\bibitem{ref12}H. Kaplan, Phys. Rev. {\bf 85}, 1038 (1952). 





\bibitem{ref11}E. P. Wohlfarth, Rev. Mod. Phys. {\bf 25}, 211 (1953).

\bibitem{ref10}J. C. Slater, Rev. Mod. Phys. {\bf 25}, 199 (1953).



\bibitem{ref9}A. H. Mitchell, Phys. Rev. {\bf 105}, 1439 (1957).


\bibitem{ref8} D. A. Goodings, Phys. Rev. {\bf 127}, 1532 (1962).





\bibitem{ref7} C. Herring, Rev. Mod. Phys. {\bf 34}, 631 (1962).


\bibitem{ref6}P.-O. Lowdin, Rev. Mod. Phys. {\bf 34}, 80 (1962).




\bibitem{ref5}R. E. Watson and A. J. Freeman, Phys. Rev. {\bf 152},
  566 (1966).



\bibitem{ref4}G. M. Copland and  P. M. Levy, Phys. Rev. B {\bf 1},
  3043 (1970).




\bibitem{ref3}F. Milstein and L. B. Robinson, Phys. Rev. Lett. {\bf
  18}, 308 (1967).



\bibitem{ref2} R. M. White and R. L. White, Phys. Rev. Lett. {\bf 20},
  62 (1968).

\bibitem{ref1} W. E. Evenson, J. R. Schrieffer and S. Q. Wang,
  J. App. Phys. {\bf 41}, 1199 (1970).





\bibitem{jpc13}G. P. Zhang and T. F. George, J. Phys.: Condens. Matter
  {\bf 25}, 366002 (2013).


\bibitem{eric}E. Beaurepaire, J. -C. Merle, A. Daunois, and
  J.-Y. Bigot, 
  \prl {\bf 76}, 4250 (1996).

\bibitem{prl00} G. P. Zhang and W. H\"ubner,  \prl {\bf 85},
  3025 (2000).


\bibitem{ourreview}G. P. Zhang \ete, Topics Appl. Phys.  {\bf 83}, 245
  (2002).


\bibitem{np09}G. P. Zhang, W. H\"ubner, G. Lefkidis, Y. Bai, and
  T. F. George,  Nature Phys.  {\bf 5}, 499
  (2009).



\bibitem{rasingreview}A. Kirilyuk, A. V. Kimel and Th. Rasing, 
  Rev. Mod. Phys. {\bf 82}, 2731 (2010).



\bibitem{durr} H. A. D\"urr, 
  Nuclear Instruments and Methods in Physics Research A {\bf 601},
  132 (2009).

\bibitem{durr1} H. S. Rhie, H. A. D\"urr and W. Eberhardt, \prl
  {\bf 90}, 247201 (2003).



\bibitem{lis}M. Lisowski, P. A. Loukakos, A. Melnikov, I. Radu,
  L. Ungureanu, M. Wolf, and U. Bovensiepen,  \prl {\bf 95}, 137402 (2005).




\bibitem{weinelt} R. Carley \ete, \prl {\bf 109}, 057401 (2012).





\bibitem{mat1}M. Matsubara, Y. Okimoto, T. Ogasawara, Y. Tomioka,
  H. Okamoto, and Y. Tokura,  Phys. Rev. Lett. {\bf 99}, 207401 (2007).

\bibitem{mat2} M.  Matsubara, A.  Schroer, A.  Schmehl, A.  Melville,
  C.  Becher, M.  T.  Martinez, D.  G. Schlom, J.  Mannhart, J.
  Kroha, and M. Fiebig,  arXiv:1304.2509 (2013).




\bibitem{bigot09} J.-Y Bigot, M. Vomir and E. Beaurepaire, Nature
  Phys. {\bf 5}, 515 (2009).


\bibitem{sz}A. Szilva \ete, \prl {\bf 111}, 127204 (2013). 


\bibitem{prb98} W. H\"ubner and G. P. Zhang, \prb {\bf 58}, R5920
  (1998).

\bibitem{mathias} S. Mathias \ete, PNAS {\bf 109}, 4792 (2012).


\bibitem{stamenova} M. Stamenova and S. Sanvito, Phys. Rev. B {\bf 88},
  104423 (2013).



\bibitem{htc}S. G. Han, Z. V. Vardeny, K. S. Wong, O. G. Symko, and
  G. Koren,  \prl {\bf 65}, 2708 (1990).

\bibitem{htc1} C. L. Smallwood, J. P. Hinton,
  C. Jozwiak, W. Zhang, J.  D. Koralek, H. Eisaki, D.-H. Lee,
  J. Orenstein, and A. Lanzara,
  Science {\bf 336}, 1137 (2012).

\bibitem{htc2} J. Graf, C. Jozwiak, C. L. Smallwood, H. Eisaki,
  R. A. Kaindl, D-H. Lee, and A. Lanzara, 
Nature Physics {\bf 7}, 805 (2011).



\bibitem{htc3} M. Beck \ete,  Phys. Rev. Lett. {\bf 107},
  177007 (2011).

\bibitem{htc4} D. N. Basov, Richard D. Averitt, D. van der Marel,
  M. Dressel, and K. Haule,  Rev. Mod. Phys. {\bf 83}, 471 (2011).
\bibitem{htc5} J. Qi \ete,  Phys. Rev. Lett. {\bf 111},
  057402 (2013).



\bibitem{stanciu}C. D. Stanciu, F. Hansteen, A. V. Kimel, A. Kirilyuk,
  A. Tsukamoto, A. Itoh, and Th. Rasing,  \prl {\bf 99}, 047601 (2007).




\bibitem{deric}DERIC program: Diatomic integrals over Slater-type
  orbitals, Stan Hagstrom, Indiana University, (2012).


\bibitem{rue} K. Ruedenberg,  J.  Chem.  Phys.  {\bf 19}, 1459 (1951).


\bibitem{freeman}A. J. Freeman, R. K. Nesbet and R. E. Watson, 
  Phys. Rev. {\bf 125}, 1978 (1962).



\bibitem{sm}R. Stuart and W. Marshall, Phys. Rev. {\bf 120}, 353 (1960).



\bibitem{prb09} G. P. Zhang, Y. Bai, and T. F. George, \prb {\bf 80},
  214415 (2009).


\bibitem{prl08} G. P. Zhang,  \prl
  {\bf 101}, 187203 (2008).


\bibitem{wein} S. Wienholdt, D. Hinzke, K. Carva, P. M. Oppeneer and
  U. Nowak, \prb {\bf 88}, 020406 (2013). U. Atxitia and
  O. Chubykalo-Fesenko, \prb {\bf 84}, 144414 (2011). 



\bibitem{ostler}T. A. Ostler \ete,
  Nat. Commun. {\bf 3}, 666 (2012).



\bibitem{mw}N. D. Mermin and H. Wagner, Phys. Rev. Lett. {\bf 17}, 1133
  (1966).











\bibitem{jpc11}G. P. Zhang,  J. Phys.: Condens. Matter {\bf 23}, 206005 (2011).



\bibitem{spin} B. Hillebrands and K. Ounadjela, {\it Spin Dynamics in
  Confined Magnetic Structures II}, Springer (2003).

\bibitem{oomf} M. Donahue and D. Porter, {\it The Object Oriented
  MicroMagnetic Framework (OOMMF) project at ITL/NIST},
  http://math.nist.gov/oommf/


\bibitem{koopmans}B. Koopmans \ete, 
  Nat. Mater. {\bf 9}, 259 (2010).


\bibitem{la}C. La-O-Vorakiat \ete,  Phys. Rev. X {\bf 2}, 011005 (2012).




\bibitem{kryder}H.-P. D. Shieh and Mark H. Kryder, IEEE Transactions
  on magnetics, {\bf 23}, 171 (1987); H. P. Shieh and M. H. Kryder,
  Appl. Phys. Lett. {\bf 49}, 473, 1986)

\bibitem{note1}The configuration number is only for our programming
  purpose to distinguish various configuration. It has no special
  meaning. 


\bibitem{lefkidismmm}G. Lefkidis and W. H\"ubner, 
  J. Mag. Mag. Mater. {\bf 321}, 979 (2009).





\bibitem{noslow}B. Vodungbo \ete,  Nature Comm. {\bf 3}, 999 (2012)




\bibitem{alebrand} S. Alebrand, A. Hassdenteufel, D. Steil,
  M. Cinchetti and M. Aeschlimann,  \prb {\bf 85},
  092401 (2012).




\bibitem{hohlfeld} J. Hohlfeld, C. D. Stanciu and A. Rebei, 
  Appl. Phys. Lett. {\bf 94}, 152504 (2009).



\bibitem{khor} A. R. Khorsand \ete,  \prl {\bf 108}, 127205
  (2012).
%





\bibitem{vah1} K. Vahaplar, A. M. Kalashnikova, A. V. Kimel,
  D. Hinzke, U. Nowak, R. Chantrell, A. Tsukamoto, A. Itoh,
  A. Kirilyuk, and Th. Rasing, \prl
  {\bf 103}, 117201 (2009).


%


\bibitem{fan}D. Steiauf and M. F\"ahnle,  \prb {\bf
  79}, 140401 (2009).

\bibitem{fan1} M.  F\"ahnle and C. Illg,  J. Phys.:
  Condens. Matter {\bf 23}, 493201 (2011).


\bibitem{elliott}R. J. Elliott,
  Phys. Rev. {\bf 96}, 266 (1954).

\bibitem{yafet} Y. Yafet, Solid State Phys. {\bf 14}, 1 (1963).


\bibitem{wien} P. Blaha, K. Schwarz, G. K. H. Madsen, D. Kvasnicka,
  and J. Luitz, WIEN2k, An augmented plane wave + local orbitals
  program for calculating crystal properties (Karlheinz Schwarz,
  Techn. Universit\"at Wien, Austria, 2001).










\bibitem{sch}S.  Essert and H.  C.  Schneider,  J. Appl. Phys. {\bf 111},
  07C514 (2012).

%



\bibitem{radu}I. Radu \ete,  Nature {\bf 472}, 205 (2011).


%


































































































































































%






























































\end{thebibliography}
\end{document}